\documentclass[twocolumn,prb,reprint,showpacs,preprintnumbers,amsmath,amssymb,groupedaddress]{revtex4-1}
\usepackage{graphicx}
\usepackage{dcolumn}
\usepackage{bm}
\usepackage{graphicx}
\usepackage{subfigure}
\usepackage{physics}

\begin{document}

\title{Thermoelectric study of dissipative quantum dot heat engines}

\author{Bitan De}
\author{Bhaskaran Muralidharan}%
 \email{bm@ee.iitb.ac.in}
\affiliation{Department of Electrical Engineering, Indian Institute of Technology Bombay, Powai, Mumbai-400076, India
}%




\date{\today}

\begin{abstract}
This paper examines the thermoelectric response of a dissipative quantum dot heat engine based on the Anderson-Holstein model in two relevant operating limits, (i) when the dot phonon modes are out of equilibrium, and (ii) when the dot phonon modes are strongly coupled to a heat bath. In the first case, a detailed analysis of the physics related to the interplay between the quantum dot level quantization, the on-site Coulomb interaction and the electron-phonon coupling on the thermoelectric performance reveals that an n-type heat engine performs better than a p-type heat engine. In the second case, with the aid of the dot temperature estimated by incorporating a {\it{thermometer bath}}, it is shown that the dot temperature deviates from the bath temperature as electron-phonon interaction in the dot becomes stronger. Consequently, it is demonstrated that the dot temperature controls the direction of phonon heat currents, thereby influencing the thermoelectric performance. Finally, the conditions on the maximum efficiency with varying phonon couplings between the dot and all the other macroscopic bodies are analyzed in order to reveal the nature of the optimum junction.   

\end{abstract}
\maketitle


\section{\label{sec:level1}Introduction}
Research in the area of nanoscale thermoelectrics is primarily carried out along two broad directions: (i) materials and devices for practical energy conversion, and (ii) fundamental studies on heat flow in the nanoscale. The former concerns materials and interface design, device optimization and integration at the systems level \cite{shakouri}. The physics that is actively considered at the materials level is that of maximizing the thermoelectric figure of merit, $zT$ \cite{Hicks1993,Hickss1993,Dresselhaus2007,Poudel2008,Snyder2008,Heremans2008,Murphy2008}, via detailed electronic structure and interface considerations. The latter is exploratory in nature and concerns fundamental transport studies on the physics of heat flow at the nanoscale \cite{Andreev2001,Humphrey2002,Kubala2006,Kubala2008,nak,Kim2014,Reddy1568,jordan1,jordan2,Agarwal2014,Sothmann2014}. It is also well known from the thermodynamic interpretation of thermoelectric processes that $zT$ is conceptually meaningful only in the linear response regime \cite{ioffe} and does not provide the full picture of heat flow at the nanoscale\cite{Muralidharan2012,jordan1,Sothmann2014,whitney}. Therefore, themoelectric analysis based on power and efficiency considerations \cite{whitney,Muralidharan2012,Sothmann2014} have gained precedence when it comes to fundamental studies \cite{nak,jordan2,Muralidharan2012,Agarwal2014,Zimb2016,Leijnse2010}.\\
\indent Zero-dimensional systems such as molecules or quantum dots are known to possess unique thermoelectric properties \cite{Mahan1996,Zimb2016} owing to their highly distorted electronic density of states (DOS). From a fundamental stand point, nanoscale heat engines are built using quantum dots or molecules sandwiched between two contacts. Thermoelectric transport measurements are performed by subjecting electrochemical potential and thermal gradients across them \cite{Kim2014,Reddy1568}.  Specific to short molecules and quantum dots, a lot of recent research work has focused on how strong correlation effects related to Coulomb charging may influence the thermoelectric performance \cite{jordan1,jordan2,Sothmann2014,Muralidharan2012,Zimb2016} under various bias situations.\\
\begin{figure*}[]
	\begin{center}
		\subfigure[]{\includegraphics[width=0.45\textwidth, height=0.25\textwidth]{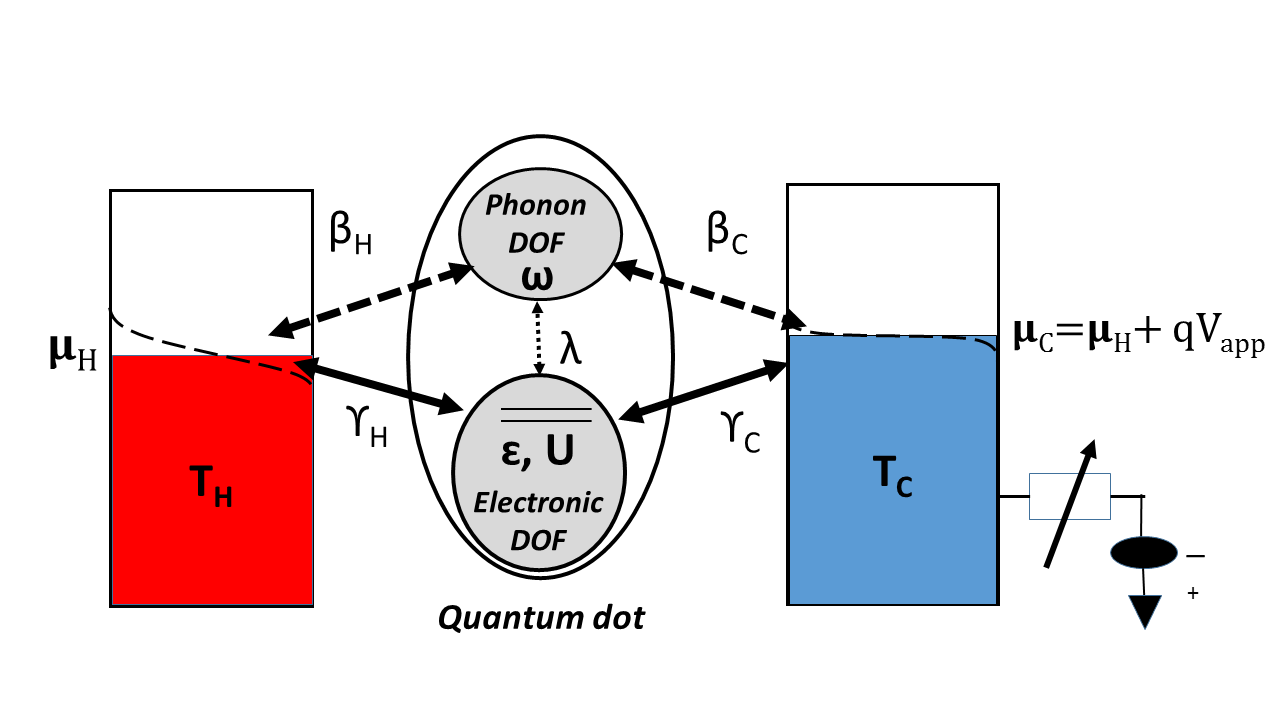}\label{1a}}
		\quad
		\subfigure[]{\includegraphics[width=0.45\textwidth, height=0.25\textwidth]{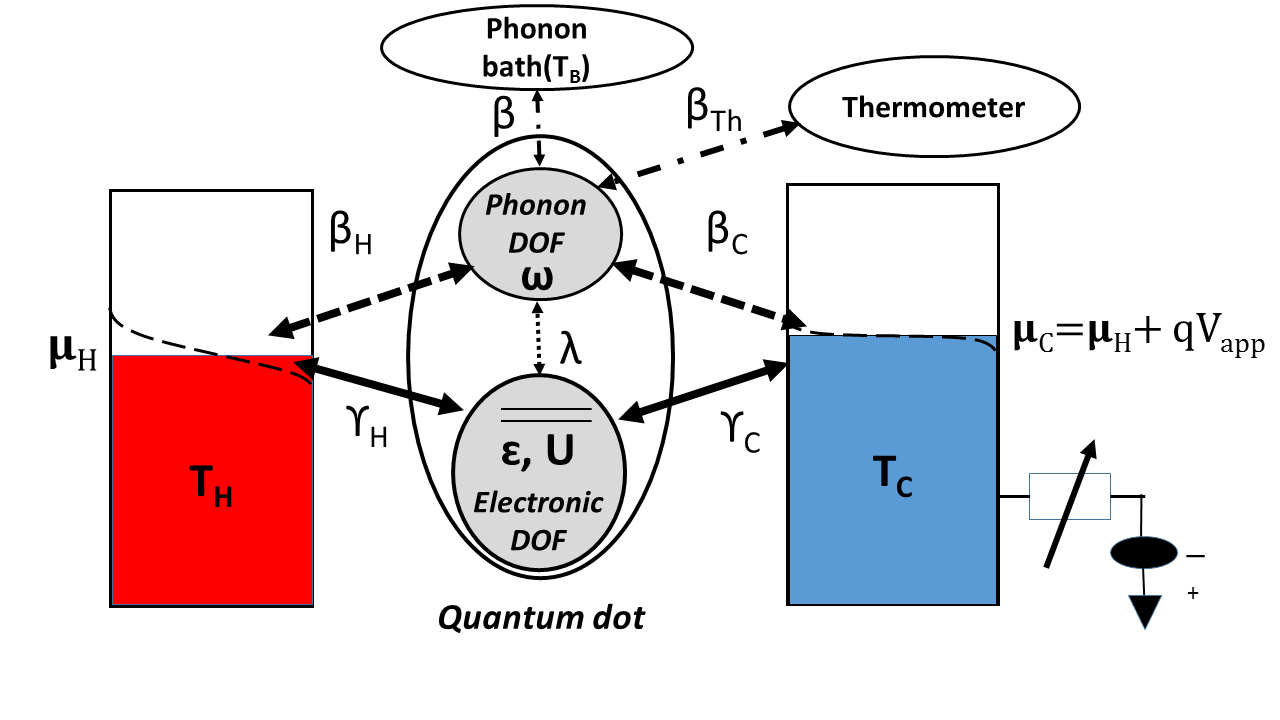}\label{1b}}
		\quad
		\subfigure[]{\includegraphics[width=0.45\textwidth, height=0.25\textwidth]{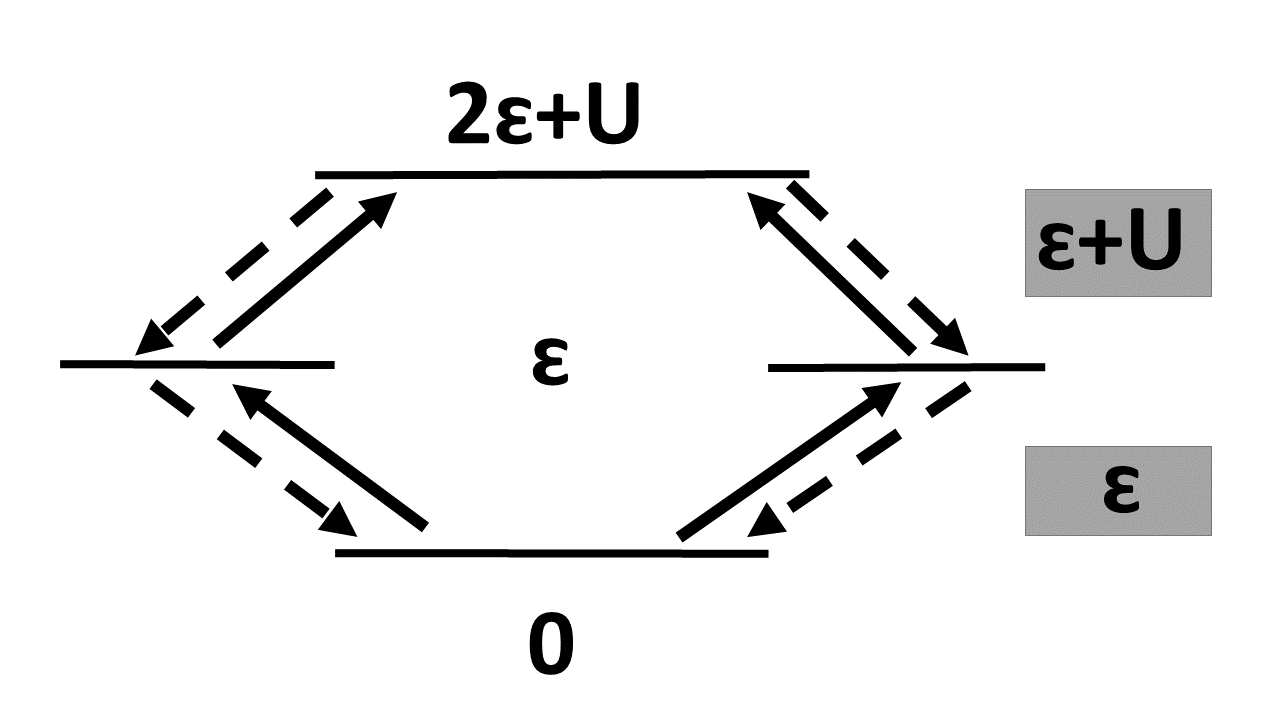}\label{1c}}
		\quad
		\subfigure[]{\includegraphics[width=0.45\textwidth, height=0.25\textwidth]{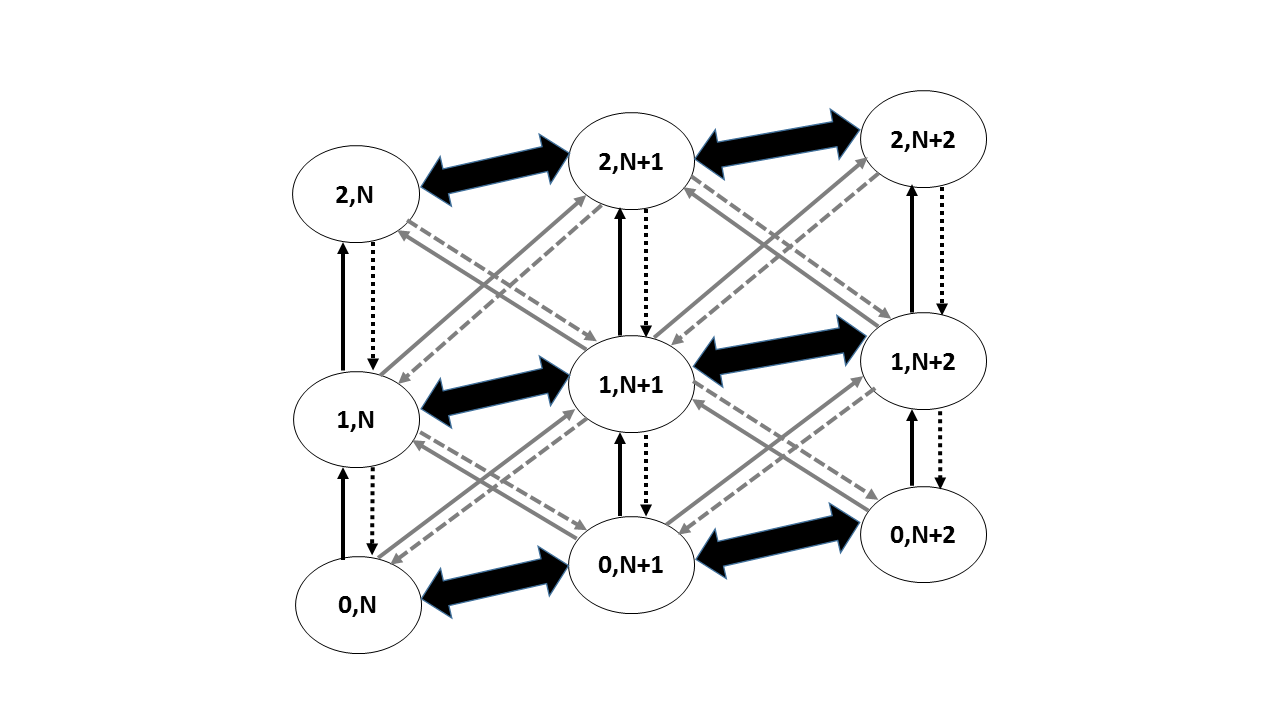}\label{1d}}
		
	\end{center}
	
	\caption{Two operating limits of the dissipative heat engine set up studied here: (a) the dot phonon degrees of freedom (DOF) are out of equilibrium, and (b) the dot phonon degrees of freedom (DOF) relax via coupling to a heat bath with a rate $\beta$. In this case, the dot temperature is estimated by coupling a {\it{thermometer bath}} weakly, with an associated rate $\beta_{th}$. In both cases, the quantum dot is  coupled weakly to the contacts with electronic rates, $\gamma_H$, $\gamma_C$ and phonon rates $\beta_H$, $\beta_C$. (c) State transition diagram in the quantum dot electronic Fock space. Electron transitions take place between the states where electron numbers differ by $\pm$ 1. (d) The state transition diagram in the electron-phonon Fock space.The black solid and black dash arrows represent electronic tunneling processes. The gray solid arrows and gray dotted arrows represent the phonon-assisted tunneling processes. The black double-sided arrows represent the heat bath assisted phonon transitions. In both cases, the currents are driven by a temperature gradient applied between the two contacts, $H$ and $C$, in a voltage controlled set up (see text). }
\end{figure*}
\indent However, the charging of the system due to electronic transport processes changes the nuclear geometry and couples with various vibrational modes \cite{Hartle2011,Zazunov2006}. The interplay between electronic and vibrational degrees of freedom have found numerous signatures in a multitude of charge transport experiments \cite{Park2000,Leroy2004,LeRoy2005,Yu2004,Sapmaz2006,Zhitenev2002}.  From the point of view of thermoelectrics, the electron-phonon interactions are important since they modify both charge and heat currents at the nanoscale. While charge and heat transport in the dissipative quantum dot set up described using the Anderson-Holstein model have been the subject of many works \cite{Siddiqui2006,Entin-Wohlman2010,Entin-Wohlman2012,Entin-Wohlman2014,Hartle2011,Zazunov2006,Segal2006,Segal2005,Segall2005}, analysis of thermoelectric transport in this regime has not received much attention \cite{Leijnse2010}. \\
\indent The object of this paper is to advance the basic understanding developed in an earlier work \cite{Leijnse2010} on dissipative quantum dot heat engines by elucidating some relevant and novel physics that arises under two experimentally relevant operating limits. The first limit, being when the dot phonon modes are in non-equilibrium. Such a situation occurs in quantum dot systems that are suspended \cite{LeRoy2005} over metallic contacts and can hence be driven out of equilibrium, giving rise to interesting charge transport signatures \cite{Siddiqui2006}. The second limit occurs when the dot phonons relax via coupling to a heat bath held at a fixed temperature \cite{Siddiqui2006,Entin-Wohlman2010,Entin-Wohlman2012,Entin-Wohlman2014}. In the first case, a detailed analysis of the physics related to the interplay between the quantum dot level quantization, the on-site Coulomb interaction and the electron phonon coupling on the thermoelectric performance is carried out. Importantly, it is demonstrated that due to such an interplay, an n-type heat engine performs better than a p-type heat engine. In the second case, with the aid of the dot temperature estimated by incorporating a {\it{thermometer bath}}, it is shown that the dot temperature deviates strongly from the bath temperature as electron-phonon interaction in the dot becomes stronger. An interesting consequence of which is that the dot temperature intimately controls the direction of phonon heat current thereby influencing the thermoelectric performance. Finally, we evaluate the trend of the maximum efficiency as the phonon couplings between the dot and all the other macroscopic bodies vary.  \\
\indent This paper is organized as follows: Sec. II introduces the Anderson-Holstein based dissipative heat engine and formulates the transport equations related to the two operating limits we consider. In Sec. IIIA and Sec. IIIB we analyze the performance of heat engines when phonons are out of equilibrium. In Sec. IIIC and Sec. IIID, we elucidate the physics related to heat engines coupled to heat baths with the aid of the dot temperature. In Sec. IV, we summarize our result and conclude.


\section{\label{sec:level2}Physics and Formulation}
\indent A schematic of the heat engine setups studied here is presented in Fig.~\ref{1a} and Fig.~\ref{1b}. Both setups constitute a quantum dot described by the dissipative Anderson Holstein Hamiltonian coupled weakly to two macroscopic contacts denoted as $H$ and $C$, which drive charge and heat currents through the dot. Additionally, the dot can be coupled strongly to a heat bath $B$, and be weakly coupled to the thermometer bath \cite{Segall2005} as presented in Fig.~\ref{1b}. We note that the set ups described here are typically {\it{voltage controlled}}, primarily driven by  the application of a temperature gradient accompanied by the control of the voltage drop via a variable load resistor.
\subsubsection{Model Hamiltonian}
\indent The composite Hamiltonian of the set up is given as $\hat{H}=\hat{H}_{D}+\hat{H}_{C}+\hat{H}_{B}+\hat{H}_{CD}+\hat{H}_{BD}$, where $\hat{H}_{D}$, $\hat{H}_{C}$, and $\hat{H}_{B}$ are the respective Hamiltonians of the dot, the contacts, and the bath, while $\hat{H}_{CD}$ and $\hat{H}_{BD}$ represent the coupling Hamiltonians between the dot and the contacts and between the dot and the bath respectively. The dot Hamiltonian is described via the Anderson-Holstein model given by
\begin{equation}
	\begin{split}
	\hat{H}_{D}=(\sum_{\sigma}^{} \epsilon_{\sigma} \hat{n}_{\sigma}+U \hat{n}_{\uparrow}\hat{n}_{\downarrow}) 
	+\hbar \omega_{\nu}\hat{n}_{\nu} \\+\sum_{\sigma}^{} \lambda_{\nu} \hbar \omega_{\nu} \hat{n}_{\sigma} (\hat{b}_{\nu}^{\dagger}+\hat{b}_{\nu}),
	\end{split}
	\label{Anderson}
\end{equation}
where the dot comprises a single spin degenerate energy level with an on-site energy, $\epsilon_{\sigma}$, and a Coulomb interaction energy, $U$. The phonon degree of freedom is described via a single phonon mode of angular frequency $\omega_{\nu}$. Inside the dot, the electrons and phonons interact via the electron-phonon coupling, $\lambda_{\nu}$. Here, $\hat{n}_{\sigma}$=$\hat{d}_{\sigma}^{\dagger} \hat{d}_{\sigma}$ and $\hat{n}_{\nu}$=$\hat{b}_{\nu}^{\dagger} \hat{b}_{\nu}$ are the dot electron and dot phonon number operators respectively, given that $\hat{d}_{\sigma}^{\dagger} (\hat{d}_{\sigma})$ and $\hat{b}_{\nu}^{\dagger} (\hat{b}_{\nu})$ represent the creation (annihilation) operator for the electrons and phonons in the dot respectively. \\

The contact and heat bath Hamiltonians and their respective coupling Hamiltonians with the dot are defined as
\begin{equation}
	\begin{split}
		\hat{H}_{C}=\sum_{\alpha\in H,C}^{} \sum_{k\sigma'}^{}\epsilon_{\alpha k\sigma'} \hat{n}_{\alpha k\sigma'}+\sum_{\alpha\in H,C}^{} \sum_{p}^{}\hbar \omega_{\alpha p} \hat{n}_{\alpha p}
	\end{split}
\end{equation}
\begin{equation}
	\begin{split}
		\hat{H}_{B}=\sum_{t}^{} \hbar \omega_{t}\hat{B}_{t}^{\dagger} \hat{B}_{t}
	\end{split}
\end{equation}
\begin{equation}
	\begin{split}
		\hat{H}_{CD}=\sum_{k\sigma', \sigma}^{} [\tau_{ \alpha k\sigma'\sigma}^{el} \hat{c}_{\alpha k \sigma'}^{\dagger}\hat{d}_{\sigma}+h.c] \\ +\sum_{\nu,p}^{}[\tau_{ \alpha p\nu}^{ph} (\hat{a}_{\alpha p}^{\dagger}+\hat{a}_{\alpha p})(\hat{b}_{\nu}^{\dagger}+\hat{b}_{\nu})+h.c]
	\end{split}
\end{equation} 
\begin{equation}
	\begin{split}
		\hat{H}_{BD}=\sum_{\nu,t}^{}\tau^{ph}_{ t\nu}(\hat{B}_{t}^{\dagger}+\hat{B}_{t})(\hat{b}_{\nu}^{\dagger}+\hat{b}_{\nu}).
	\end{split}
\end{equation}
Here, $\hat{n}_{\alpha k \sigma'}$=$\hat{c}_{\alpha k\sigma'}^{\dagger} \hat{c}_{\alpha k\sigma'}$ and  $\hat{n}_{\alpha p}$=$\hat{a}_{\alpha p}^{\dagger} \hat{a}_{\alpha p}$ are the electron and phonon number operators in the contacts. The contacts are assumed to be in the eigen-basis with wave vectors $k$ and spin orientation ${\sigma'}$. An electron in the dot with a spin orientation $\sigma$ is coupled to an electron in contact $\alpha$ ($\alpha \in H,C$) through $\tau_{ \alpha k\sigma'\sigma}^{el}$. Similarly $\nu$-th phonon mode of the dot is coupled to the $t$-th phonon mode of the contact through $\tau_{ t\nu}^{ph}$.\\
\indent The dot Hamiltonian $\hat{H}_{D}$ is diagonalized by the polaron transformation \cite{Braig2003,Siddiqui2006} leading to the renormalization of the on-site and Coulomb interaction energies given by
\begin{equation}
\tilde{\epsilon}_{\sigma}=\epsilon_{\sigma}-(\lambda^2 \hbar \omega_{\nu}),
\end{equation}
\begin{equation}
\tilde{U}=U-(2\lambda^2 \hbar \omega_{\nu}).
\end{equation}
The renormalized dot many-particle energies are given by $E=\tilde{E}_{\sigma}+m \hbar \omega_{\nu}$, where $m=(0,1,2,3,...)$ and $\tilde{E}_{\sigma}$=$(0,\tilde{\epsilon}_{\uparrow},\tilde{\epsilon}_{\downarrow},\tilde{\epsilon}_{\uparrow}+\tilde{\epsilon}_{\downarrow}+\tilde{U})$.\\
 \indent Both $\hat{H}_C$ and  $\hat{H}_B$ remain unchanged due to the renormalization since they are independent of the dot operators. The transformation of the electron tunneling part of the Hamiltonian $\hat{H_{CD}}$ leads to a modification of the electron coupling factor, $\tilde{\tau}_{ \alpha k\sigma'\sigma}^{el}=\tau_{ \alpha k\sigma'\sigma}^{el}\exp[-\lambda_{\nu}(\hat{b_{\nu}}-\hat{b_{\nu}}^{\dagger})]$. We can neglect the renormalization of phonon coupling factors $\tau_{ \alpha p\nu}^{ph}$ and $\tau^{ph}_{ t\nu}$,  considering that both of them are very small, which is an essential condition to get optimized thermoelectric efficiency \cite{Leijnse2010}. \\
 \indent With the above definitions, in the calculations to follow, it is also customary to define various tunneling rates under the assumption of dispersionless contacts as follows: The electronic tunneling rate between the dot and the contact $\alpha$ with density of states $\rho_{\alpha}$ is derived from the Fermi's golden Rule as  $\gamma_{\alpha}=\frac{2\pi}{\hbar}\sum_{\sigma} |\tilde{\tau}^{el}_{\alpha k\sigma}|^2 \rho_{\alpha \sigma}$. Similarly, the phonon relaxation rates between the dot and other macroscopic bodies $r'$( where, $r' \in H,C,B$) with phonon density of states $D_{r'}$ is expressed as $\beta_{r'}=\frac{2\pi}{\hbar}|\tau^{ph}_{r'\nu}|^2D_{r'}$. 
\subsubsection{Transport formulation}
We first state the important assumptions made in the set ups that we consider. First, we work in a regime where the dot-bath phonon relaxation rate is smaller compared to the tunneling rates between the dot and the contacts. Larger phonon couplings may cause further energy shift in the dot phonon modes leading to a non-separable terminal phonon currents \cite{Segal2006,Segal2005,Segall2005}. The assumption of small phonon coupling also allows us to exclude system damping \cite{Braig2003}. Second, we perform all calculations within the sequential tunneling limit \cite{Timm,Leijnse2010}, where, the associated tunneling energies, $\hbar \gamma_{\alpha}, \hbar \beta_{r} <<  k_BT$. The mathematical expressions for the electron tunneling rate $\gamma$ and the phonon relaxation rate $\beta$ are to be defined shortly. The sequential tunneling limit is the relevant regime when describing quantum dot transport as most experiments are performed in this regime \cite{tarucha1,Timm}. Under this approximation, given a spin degenerate level coupled to non magnetic contacts, transport is described via rate equations \cite{Beenakker,Basky_Beenakker,Basky_Datta,Timm} in the diagonal subspace of the quantum dot reduced density matrix \cite{Koenig_1,Koenig_2,Brouw,Milena_noncoll,Basky_Milena}. \\
\indent The use of the diagonal subspace is justified in the absence of coherences. In the current context, we are faced with two types of coherences, (a) coherence between the degenerate up-spin and down-spin levels and (b) coherences between various phonon induced side band energies. The first type can be neglected simply because electron-phonon interaction is described via a coupling factor $\lambda_{\nu}$, which is spin independent. Such a coherence between up-spin and down-spin levels is characteristic of systems with non-collinear magnetic contacts \cite{Koenig_1,Koenig_2,Basky_Milena}, or in systems which have orbital degeneracies \cite{Brouw,Milena_noncoll}. The second type, namely, the coherence between two phonon induced side bands can be safely neglected by assuming that the energy spacing between two adjacent side bands is larger than the tunneling induced broadening of energy levels, i.e., $\hbar\omega_{\nu}>>\hbar\gamma$. In this limit electron-phonon interactions that occur at two consecutive times are completely uncorrelated \cite{Theses1}, and a Markovian approximation is also justified \cite{Timm}. This allows us to neglect bath memory also. Hence, the secular terms in density matrix get decoupled from the off-diagonal terms, which ultimately implies that such coherences may be safely ignored.  \\
\indent The electronic tunneling rate between two electron-phonon Fock states, $\mid n,q \rangle$ and $ \mid n \pm 1,q \rangle$, with $n$ and $q$ representing the electronic and phonon state label respectively, is given by
\begin{equation}
\begin{gathered}
R_{(n,q)\rightarrow(n+1,q')}^{el}=\\\sum\limits_{\alpha \in H,C}^{}\gamma_{\alpha}|\bra{n,q}\hat{\tilde{{d}}}_{\sigma}\ket{n+1,q'}|^2\\ \times f_{\alpha}\bigg(\frac{E_{(n+1,q')}-E_{(n,q)}-\mu_{\alpha}}{k_B T_{\alpha}}\bigg) 
\end{gathered}
\end{equation}
\begin{equation}
\begin{gathered}
R_{(n,q)\rightarrow(n -1,q')}^{el}=\\\sum\limits_{\alpha \in H,C}^{} \gamma_{\alpha}|\bra{n,q}\hat{\tilde{{d}}}_{\sigma}^{\dagger}\ket{n-1,q'}|^2\\ \times \bigg[1-f_{\alpha}\bigg(\frac{E_{(n,q)}-E_{(n-1,q')}-\mu_{\alpha}}{k_B T_{\alpha}}\bigg)\bigg]
\end{gathered}
\end{equation}
The relaxation of the dot phonons to the contacts and the heat bath cause transition between the states $(n,q)$ and $(n,q\pm 1)$ follow the Boltzmann ratio:
\begin{equation}
\begin{split}
R_{(n,q)\rightarrow(n,q+1)}^{ph}=\sum\limits_{r'\in H,C,B}^{}\beta_{r'} (q+1) exp(-\frac{\hbar \omega_{r}}{k_B T_{r'}})
\end{split}
\end{equation}
\begin{equation}
\begin{split}
R_{(n,q)\rightarrow(n,q-1)}^{ph}=\sum\limits_{r'\in H,C,B}^{}\beta_{r'} (q+1).
\end{split}
\end{equation} 
With various rates defined above, the master equation for the probabilities, $P_{n,q}$ of the many particle states $\mid n,q \rangle$, then reads:
\begin{equation}
\begin{split}
\frac{d P_{(n,q)}}{dt}=\sum_{k'=0}^{N_q'-1}\bigg[R_{(n\pm 1,q')\rightarrow(n,q)}^{el}P_{(n\pm 1,q')} \\ - R_{(n,q)\rightarrow(n\pm 1,q')}^{el}P_{(n,q)} \bigg]\\+\bigg[R_{(n ,q\pm 1)\rightarrow(n,q)}^{ph}P_{(n,q\pm 1)} \\ -R_{(n,q)\rightarrow(n,q\pm 1)}^{ph} P_{(n,q)} \bigg].
\end{split}
\end{equation}
In steady state, we set $\frac{d P_{(n,q)}}{dt}=0$, and find the null space of the rate matrix to evaluate the steady state probabilities. Using the steady state probabilities, we can get the expressions for the terminal electronic charge currents $J$ and heat currents $J_{el_{\alpha}}^{Q}$,$J_{ph_{r'}}^{Q}$  as
\begin{equation}
\begin{split}
J_{\alpha}=\sum_{n,q=0}^{N_e,N_p-1}\sum_{q'=0}^{N'_q-1} -q\bigg[R_{(n \pm 1,q') \rightarrow(n,q)}^{el_{\alpha}}P_{(n \pm 1,q')}\\ -R_{(n,q) \rightarrow(n \pm 1,q')}^{el_{\alpha}}P_{(n,q)}\bigg]
\end{split}
\end{equation}
\begin{equation}
\begin{gathered}
J_{el_{\alpha}}^{Q}=\sum_{n,q=0}^{N_e,N_p-1}\sum_{k'}^{N'_p-1} \bigg[E_{(n\pm1,q')}-\mu_{\alpha}\bigg] \\ \times R_{(n\pm 1,q') \rightarrow(n,q)}^{el_{\alpha}}P_{(n \pm 1,q')}\\ - \bigg[E_{(N_e,N_p)}-\mu_{\alpha}\bigg]\\ \times R_{(n,q) \rightarrow(n \pm1,q)}^{el_{\alpha}}P_{(n,q)}
\end{gathered}
\end{equation}
\begin{equation}
\begin{split}
J_{ph_{r'}}^{Q}=\sum_{n,q=0}^{N_e,N_p-1} \hbar \omega_{r} \bigg[R_{(n,q) \rightarrow(n,q \pm 1)}^{ph_{\alpha}}P_{(n,q)}\\- R_{(n,q \pm 1) \rightarrow(n,q)}^{ph_{\alpha}}P_{(n,q +1)}\bigg]
\end{split}
\end{equation}
The charge or electronic heat currents associated with contacts $\alpha=H(C)$ involve only the rates associated with the respective contact. However, the phonon heat current is associated with both the contacts as well as the heat bath. 
\subsubsection{Calculation of power and efficiency}
A thermal bias applied across the contacts, $T_H$, at the hot contact $H$, and $T_C$, at the cold contact $C$, can result in charge and heat currents. In the voltage controlled setup, a variable resistor controls the back flow charge current. At a voltage $V_S$, the back flow current completely cancels the charge current set up by the temperature gradient. This is referred to as the built-in potential or Seebeck voltage. The set up hence functions as a heat-to-charge-current converter or a heat engine in the voltage range $[0,V_S]$, which we term as the operating region. The electrical power generated in the circuit is given b $P=-J_{\alpha}\times V_{app}$. The thermoelectric efficiency is then expressed as
\begin{equation}
\eta=\frac{P}{J^Q_{in}},
\end{equation}
where, the input heat current includes both the electron and phonon heat currents such that the net heat input is $J^Q_{in}= J_{{el}_{H}}^{Q} + J_{{ph}_{in}}^{Q}$. It must be noted that while the input electronic heat current can be supplied only from the hot contact, the phonon heat current can be supplied from contacts or the heat bath depending on the dot temperature $T_M$. This aspect will be studied in detail in a later section.
\section{Results}
\subsection{Non-equilibrium phonons}
We first elaborate on the effect of the electron-phonon interaction parameter, $\lambda$, on the delivered electronic power, $P$, and the efficiency, $\eta$. Each operating point is signified by a constant applied thermal gradient ($T_H$=10K,$T_C$=5K) and a variable voltage bias $V$ between the two contacts. In the current section we assume that the Coulomb interaction is kept much larger, i.e., $U>>k_BT$, the electronic contact coupling and the phonon couplings, $\gamma_H,\gamma_C,\beta_H,\beta_C$ are small enough ($\hbar\gamma_{H(C)}=5\times10^{-6} eV,\hbar\beta_{H(C)}=2\times10^{-12} eV$) to keep the tunneling induced broadening of the states in the quantum dot small and ensure that the dot phonons are out of equilibrium. Transport under the sequential tunneling limit where the rate equation formalism is applicable is also ensured under these conditions.  Additionally, in this part, we consider that the dot functions as an n-type, i.e., when $\epsilon-\mu>0$.\\
\begin{center}
			\begin{figure}[!htb]
				
				\subfigure[]{\includegraphics[height=0.225\textwidth,width=0.225\textwidth]{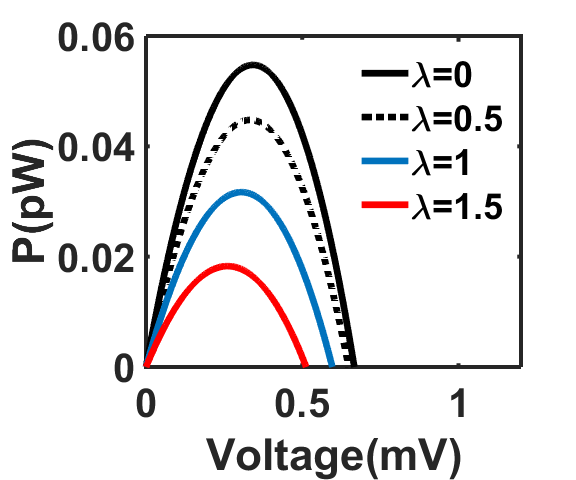}\label{2a}}
				\quad
				\subfigure[]{\includegraphics[height=0.225\textwidth,width=0.225\textwidth]{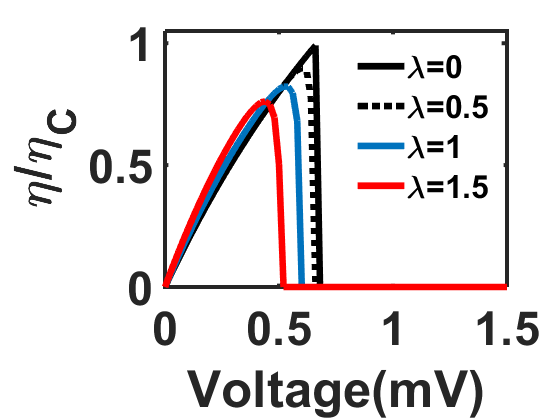}\label{2b}}
				\quad
				\subfigure[]{\includegraphics[height=0.225\textwidth,width=0.225\textwidth]{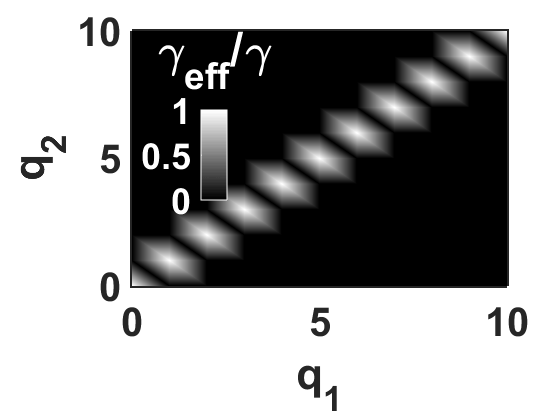}\label{2c}}
				\quad
				\subfigure[]{\includegraphics[height=0.225\textwidth,width=0.225\textwidth]{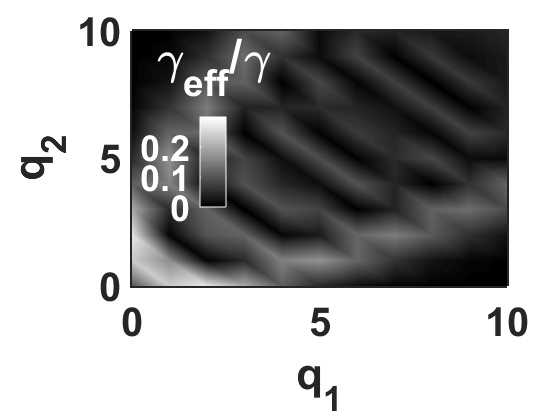}\label{2d}}
				\quad
				\subfigure[]{\includegraphics[height=0.225\textwidth,width=0.225\textwidth]{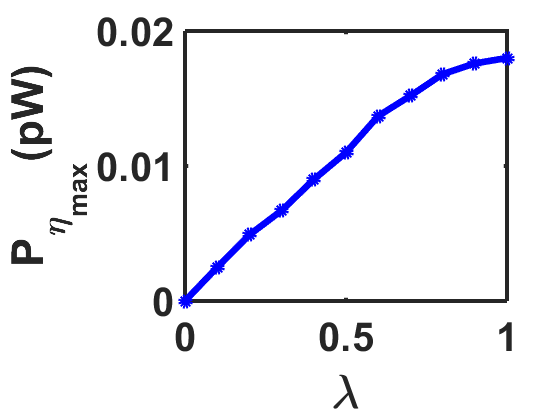}\label{2e}}
				\quad
				\subfigure[]{\includegraphics[height=0.225\textwidth,width=0.225\textwidth]{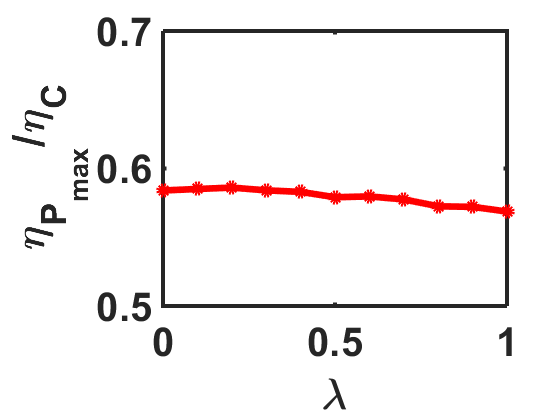}\label{2f}}
				\quad		
\caption{Thermoelectric performance with non-equilibrium phonons. The Carnot efficiency is set to $\eta_C=0.5$ at $\epsilon-\mu<2KT$. (a) Variation of the electronic power as a function of voltage as  $\lambda$ is varied. (b) Variation of $\eta$ as a function of voltage as  $\lambda$ is varied. (c) and (d) 3-D color plots of $\gamma_{eff}/\gamma$ pertaining to the tunneling between two states with phonon numbers, $q_1$ and $q_2$, for $\lambda=0$ and $\lambda=2$ respectively. As $\lambda$ is increased, we notice a larger off-diagonal contribution in the associated phonon numbers between the two states. (e) Electronic power delivered at $\eta_{max}$ as $\lambda$ increases, and (f) Variation of $\eta$  at $P_{max}$, representing the efficiency at maximum power as $\lambda$ is increased. We infer that an increase in the electron-phonon coupling leads to a better performance in terms of the power at maximum efficiency while maintaining the efficiency at maximum power. }
\end{figure}
\end{center}
\begin{center}
	\begin{figure}[!htb]
		
		\subfigure[]{\includegraphics[height=0.221\textwidth,width=0.222\textwidth]{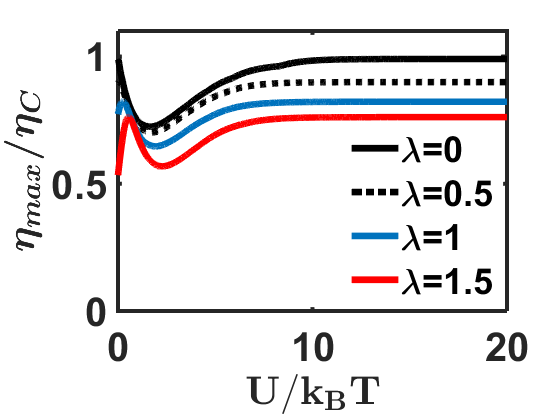}\label{3a}}
		\quad
		\subfigure[]{\includegraphics[height=0.221\textwidth,width=0.222\textwidth]{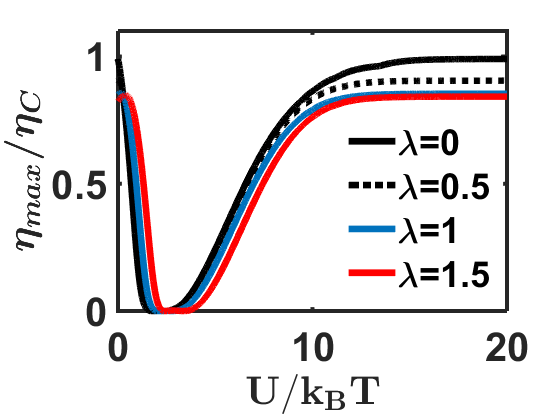}\label{3b}}
		\quad
		\subfigure[]{\includegraphics[height=0.221\textwidth,width=0.222\textwidth]{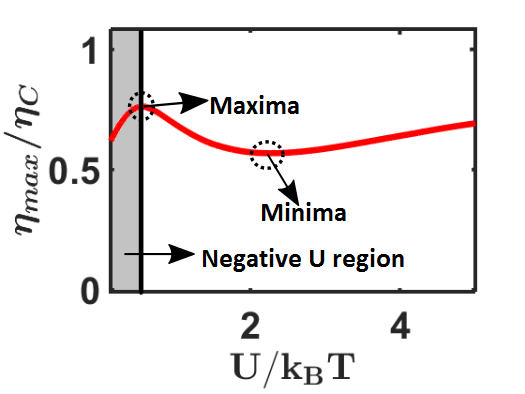}\label{3c}
		}
		\quad
		\subfigure[]{\includegraphics[height=0.221\textwidth,width=0.222\textwidth]{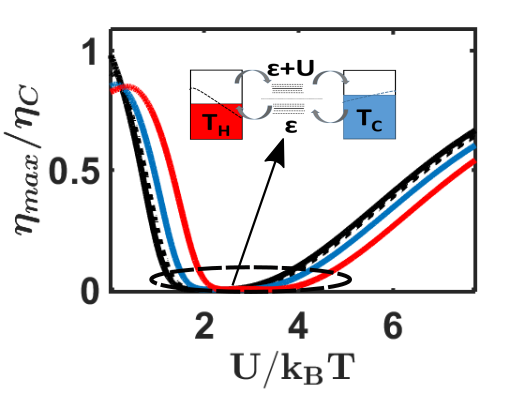}\label{3d}
		}
		\quad
		\caption{A performance comparison between n-type and p-type heat engines. Variation of $\eta_{max}$ with Coulomb interaction and electron-phonon interaction for (a) an n-type ($\epsilon-\mu=1 meV$) and (b) a p-type setup ($\epsilon-\mu=-1 meV$). For both set ups, $\eta_{max}$ maximizes at $\tilde{U}=0$. The p-type setup can have a vanishing $\eta_{max}$ at the particle-hole symmetry point. (c) Maxima and minima points of $\eta_{max}$ for an n-type engine with $\lambda=1.5$, with the grey region depicting the negative $U$ regime. (d) The region around the particle-hole symmetry point for a p-type setup where $\eta_{max}$ vanishes. We note that an n-type setup can provide a better $\eta_{max}$ in the presence of a finite $U$ since it can avoid particle-hole symmetry.}
	\end{figure}
\end{center}
\indent  A finite electron-phonon coupling causes a displacement of the potential profile of the dot and alters the electron tunneling rate between two electron-phonon states, $\ket{n,q_1}$ and $\ket{n\pm 1,q_2}$ to $\gamma_{eff}$\cite{Theses} defined as
\begin{equation}
\begin{gathered}
\gamma_{eff}=\gamma\times |C_{q_1q_2}|^2\\
=\gamma\times exp(-\frac{\lambda^2}{2})\times {\bigg(\frac{q!}{Q!}\bigg)}^2\\\times\lambda^{Q-q}\times L_{q}^{Q-q}(\lambda^2)\times[sgn(q_1-q_2)]^{(q_1-q_2)},
\end{gathered}
\end{equation}
where $q=min(q_1,q_2)$ and $Q=max(q_1,q_2)$, $C_{q_1q_2}$ is a measure of the overlap between two many body electron-phonon states with phonon numbers, $q_1$ and $q_2$, arising from the electron-phonon interaction. Referring to Fig.~\ref{2a}, we see that the peak power, as well as the Seebeck voltage, $V_S$, drops as the electron-phonon coupling parameter $\lambda$ is increased.  As $\lambda$ is increased, the charge current as well as the peak power falls, since $\gamma_{eff}$ between two states become smaller. In Fig.~\ref{2c}, we see that for $\lambda=0$, $\gamma_{eff}\sim\gamma$ is only non-zero between two states with equal phonon number. Hence, in the non-interacting case, only direct tunneling is feasible. As $\lambda$ is increased, strong electron-phonon interaction leads to the  suppression of  direct tunneling and the facilitation of phonon-assisted tunneling. \\ 
\indent We see in Fig.~\ref{2d} that a nonzero $\lambda$ results in phonon assisted tunneling between two states with unequal phonon numbers. Evidence of this phenomenon has been found in various transport experiments \cite{Steele1103,Park2000,Leturcq2008} and also has been demonstrated theoretically \cite{Koch2005,Kochh2006}. Notice that $\gamma_{eff}$ for non-zero $\lambda$ is always less than $\gamma$, due to electron-phonon coupling making the set up dissipative. For this reason, the charge current decreases and the open-circuit point is reached at a smaller voltage, leading to a fall in $V_S$ as $\lambda$ is increased. Turning to the analysis of the efficiency $\eta$, first we note the well known result \cite{Muralidharan2012} that for $\lambda=0$, $\eta$ attains a maximum of $\eta_C$ at $V_S$. But as $\lambda$ is increased, due to phonon assisted tunneling, $J^{Q}_{el}$ is always greater than $P$ within the operating range. Hence according to (16), $\eta_{max}$ falls below $\eta_C$.\\
\indent One must appreciate the fact that although a non-interacting system gives the maximum $\eta$, the power it delivers at that point is identically zero. With the inclusion of electron-phonon interaction, we evaluate two different trends: (a) Power at maximum efficiency $P_{\eta_{max}}$, and (b) Efficiency at maximum power $\eta_{P_{max}}$. In Fig.~\ref{2e} we see that the electronic power delivered at $\eta_{max}$ increases monotonically as $\lambda$ increases. On the other hand, $\eta$ at maximum power keeps almost constant as shown in Fig.~\ref{2f}. Hence, although there is a fall in the peak power as $\lambda$ increases,  $\eta_{P_{max}}$ increases slightly. This may be counted as an advantage of having stronger electron-phonon interaction. Thus, so far, we can conclude from here that when phonons are out of equilibrium, $\lambda$ is the deciding factor for the thermoelectric performance. 
\subsection{Comparison between n-type and p-type heat engines}
\begin{center}

		\begin{figure}[!htb]
			
			\subfigure[]{\includegraphics[height=0.225\textwidth,width=0.225\textwidth]{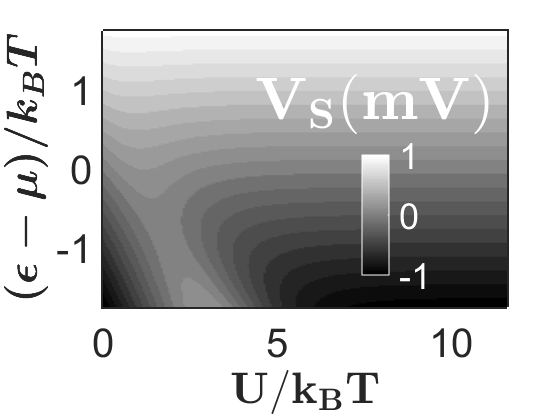}\label{4a}}
			\quad
			\subfigure[]{\includegraphics[height=0.225\textwidth,width=0.225\textwidth]{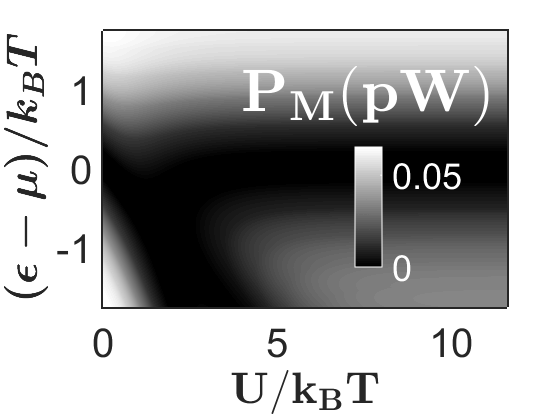}\label{4b}}
			
			\caption{Peak power characteristics of the heat engine as a function of $U$ and on-site energy position, keeping $\lambda$ constant. (a) Variation of  $V_S$ as a function of $(\epsilon-\mu)/k_{B}T$ and $U/k_{B}T$. (b) Variation in peak electronic power as a function of $(\epsilon-\mu)/k_{B}T$ and  $U/k_{B}T$. We note that $V_S$ and $P_M$ remain independent of $U$ for an n-type setup but decrease significantly for a p-type setup as the particle-hole symmetry condition is approached.}
			
		\end{figure}
		
	\end{center}
\indent We now turn our attention to an analysis with the inclusion of Coulomb interaction $U$, which brings to fore the difference between an n-type set up and a p-type set up. If $\epsilon>\mu$, then the thermal bias induced current flows from cold to hot contact, whereas the current direction is just reverse for $\epsilon<\mu$. The nomenclature due to the sense of particle flow being identical to that noted in the thermoelectric transport of n-type and p-type semiconductors . However, turning on $U$ for a p-type set up may give rise to a situation where transport channels $\epsilon$ and $\epsilon+U$ in conjunction with the phonon sidebands may give rise to a particle-hole symmetry, which will not be possible for an n-type setup. \\
\indent A schematic of the variation in $\eta_{max}$ with respect to $U$ for an n-type set up shown in Fig.~\ref{3a}. We notice that for a non-zero $\lambda$, $\eta_{max}$ never equals the Carnot efficiency and maximizes for a non-zero Coulomb interaction. In Fig.~\ref{3c}, we present a zoomed-in view for $\lambda=1.5$, which clearly shows the maxima and minima of $\eta_{max}$. We notice that $\eta_{max}$ reaches a maximum when $\tilde{U}$ disappears. Hence, electron-phonon interaction results in a region of increasing $\eta_{max}$ in the negative $\tilde{U}$ regime  \cite{Andergassen2011,Alexandrov2002}, shown as a gray shaded area in  Fig.~\ref{3c}. In Fig.~\ref{3b} we depict the $\eta_{max}$ variation for a p-type heat engine, which follows a similar trend except that it vanishes when the particle-hole symmetry point is reached \cite{Wang2010,Dubi2009,Rejec2012,Buddhiraju2015}, as shown schematically in Fig.~\ref{3d}. \\
\indent Next we compare the power generation of the n-type and the  p-type setup. In Fig.~\ref{4a} and Fig.~\ref{4b}, we detail the variation in the Seebeck voltage $V_S$ and the peak power $P_M$ as a function of  $U$ and the relative onsite-energy $\epsilon-\mu$. Both $V_S$ and $P_M$ rise with increasing $|\epsilon-\mu|$, since more voltage bias is needed to reach the open-circuit point, justifying the increase of $|V_S|$. As the operating region $[0, V_S]$, of the heat engine broadens, the peak power also increases. We see that for the n-type engine, where $\epsilon-\mu>0$, the variation of both $V_S$ and $P_M$ remains almost constant with $U$ while for a  p-type engine, where $\epsilon-\mu<0$, this is not so . 
\begin{center}
		\begin{figure}[!htb]
			
			\subfigure[]{\includegraphics[height=0.221\textwidth,width=0.229\textwidth]{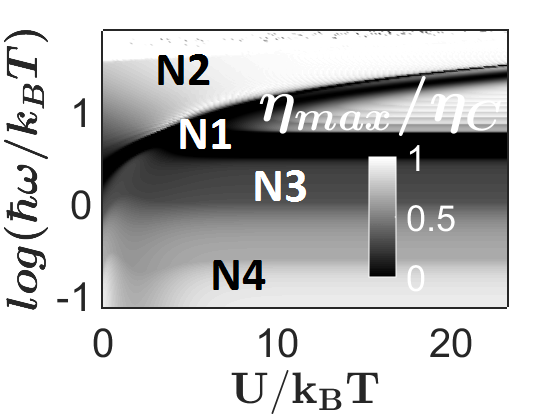}\label{5a}}
			\quad
			\subfigure[]{\includegraphics[height=0.221\textwidth,width=0.229\textwidth]{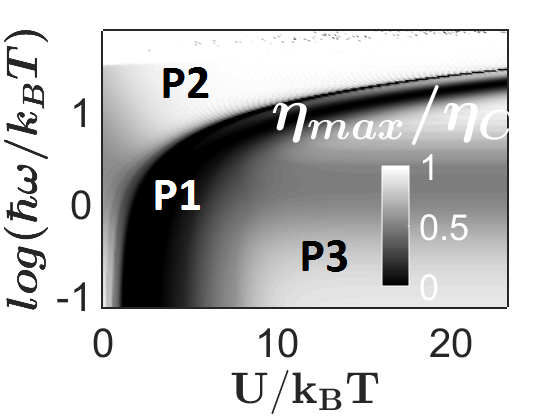}\label{5b}}
			
			\caption{Study of $\eta_{max}$ as cumulative function of $U$ and the phonon angular frequency $\omega$, keeping $\lambda$ constant ($\lambda=0.5$) for (a) an n-type and (b) a p-type set up. (b) The region $\textbf{P1}$ depicts the locus along which $\eta_{max}$ vanishes for the p-type setup. Here both $U$ and $\omega$ contribute to achieve the particle-hole symmetry. The region $\textbf{P3}$ depicts where the p-type setup can perform as a heat engine. The region $\textbf{P2}$ represents the out of the operating limit region. In (a), an n-type setup performs like a p-type setup for large $\omega$ with the regions $\textbf{N1}$ and $\textbf{N2}$ being similar to $\textbf{P1}$, $\textbf{P2}$. The region $\textbf{N3}$ has a low efficiency since $\epsilon\sim\mu$. For small $\omega$, the n-type setup achieves a high efficiency which is independent of $\omega$. Hence in the limit of low $\omega$, an n-type setup is a better heat engine. }
			
		\end{figure}
		
	\end{center}
We see that for a p-type heat engine, significant power is delivered at small values of $U$. If we increase $U$, first both $|V_S|$ and $P_M$ drops to zero before increasing again to the previous value. 
This is again due to the particle-hole symmetry condition. The analysis in Fig. 3 and Fig. 4 thus clearly indicates that an n-type engine can avoid particle-hole symmetry condition and hence performs better than the p-type engine. \\	
\indent An important aspect to be noticed is that particle-hole symmetry can be reached in two ways, either by changing $U$ or by tuning $\omega$. In Fig.~\ref{5a} and Fig.~\ref{5b} we produce a 3-D plots for the variation of $\eta_{max}$ for the n-type dot and p-type dot respectively. According to (2) and (3), for large values of $\omega$, an n-type setup performs like a p-type setup. So the upper half of Fig.~\ref{5a} (where $\omega$ is high) resembles that of Fig.~\ref{5b}. For low frequencies, we see that $\eta_{max}$ does nullify along the black branch $\textbf{N3}$, where $\tilde{\epsilon}$ almost merges with $\mu$. In the region $\textbf{N4}$, we get a high $\eta_{max}$, which is almost independent of $U$. \\
\begin{center}
	\begin{figure}[!htb]
		
		\subfigure[]{\includegraphics[height=0.221\textwidth,width=0.229\textwidth]{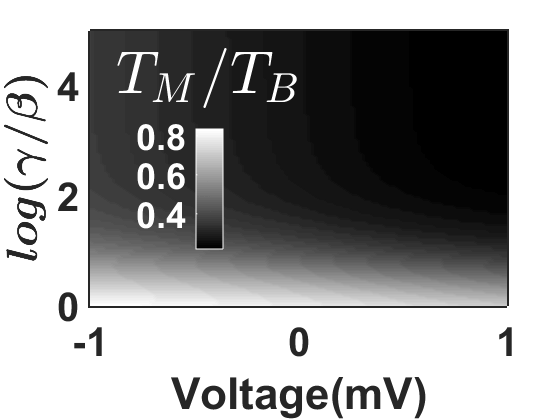}\label{6a}}
		\quad
		\subfigure[]{\includegraphics[height=0.221\textwidth,width=0.229\textwidth]{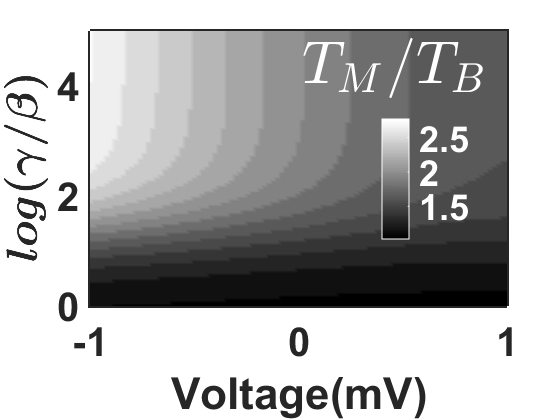}\label{6b}}
		\quad
		\subfigure[]{\includegraphics[height=0.221\textwidth,width=0.229\textwidth]{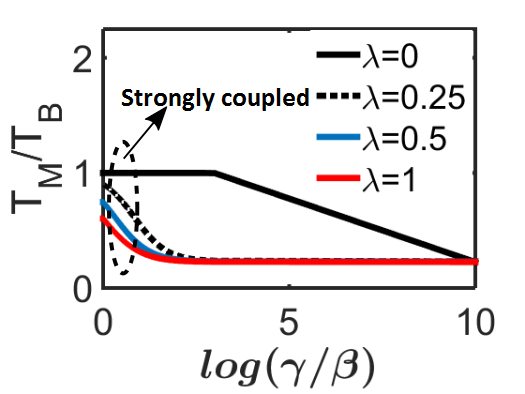}\label{6c}}
		\quad
		\subfigure[]{\includegraphics[height=0.221\textwidth,width=0.229\textwidth]{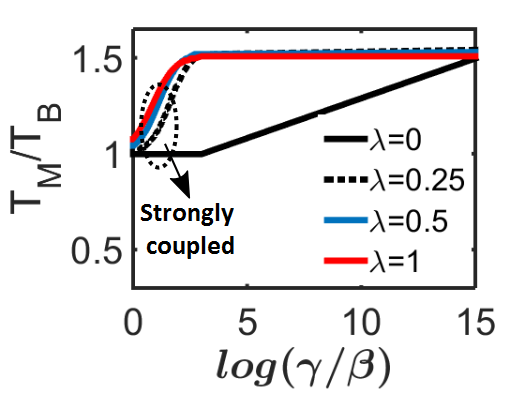}\label{6d}}
		\caption{Trends in the dot temperature $T_M$. (a),(b) Variation of $T_M$ as a function of $V_S$ and $\gamma/\beta$ for (a) $T_B>>T_H,T_C$, and (b) $T_B<<T_H, T_C$. It shows that $T_M$ is almost independent of $V_S$. (c),(d) Variation of $T_M$ as a function of $\gamma/\beta$ for different $\lambda$, for the hot and cold bath conditions respectively. In the limit of strong dot to bath coupling, $T_M$ just follows $T_B$ for $\lambda=0$, but deviates considerably as $\lambda$ increases. This implies that a bath-to-dot heat current is also feasible and will control the calculation of efficiency.}
		\end{figure}
\end{center}

\indent Switching our attention to Fig.~\ref{5b}, for the case of the p-type heat engine, we see that $\eta_{max}$ vanishes along the black branch (region marked as $\textbf{P1}$), which represents the locus of the particle-hole symmetry points. In the low frequency region marked $\textbf{P3}$, we get a comparatively high value of $\eta_{max}$ as described in Fig. 3. On the other hand, in the high frequency region (marked $\textbf{P2}$), both the polaronic shifted energy channels and their corresponding phonon sidebands go out of the transport window. Theoretically we get high efficiency in this region but it is of no use since this region lies outside the operating region.\\
\indent In general, hence, we should be interested in the low-frequency range since it serves as the power generating region, where $\eta_{max}$ is more for the n-type setup. Hence the overall study confirms that n-type engine is optimal compared to the p-type engine. One fact must be noted that we have chosen $\omega$ so that contact phonon heat current $J_{ph}^{Q}$ is low enough to control $Q^{in}$ or $\eta$. However $J_{ph}^{Q}$ becomes significant as the dot is strongly coupled to phonon modes of macroscopic bodies which we discuss in the subsequent sections. \\
\begin{center}
	\begin{figure}[!htb]
		
		\subfigure[]{\includegraphics[height=0.22\textwidth,width=0.22\textwidth]{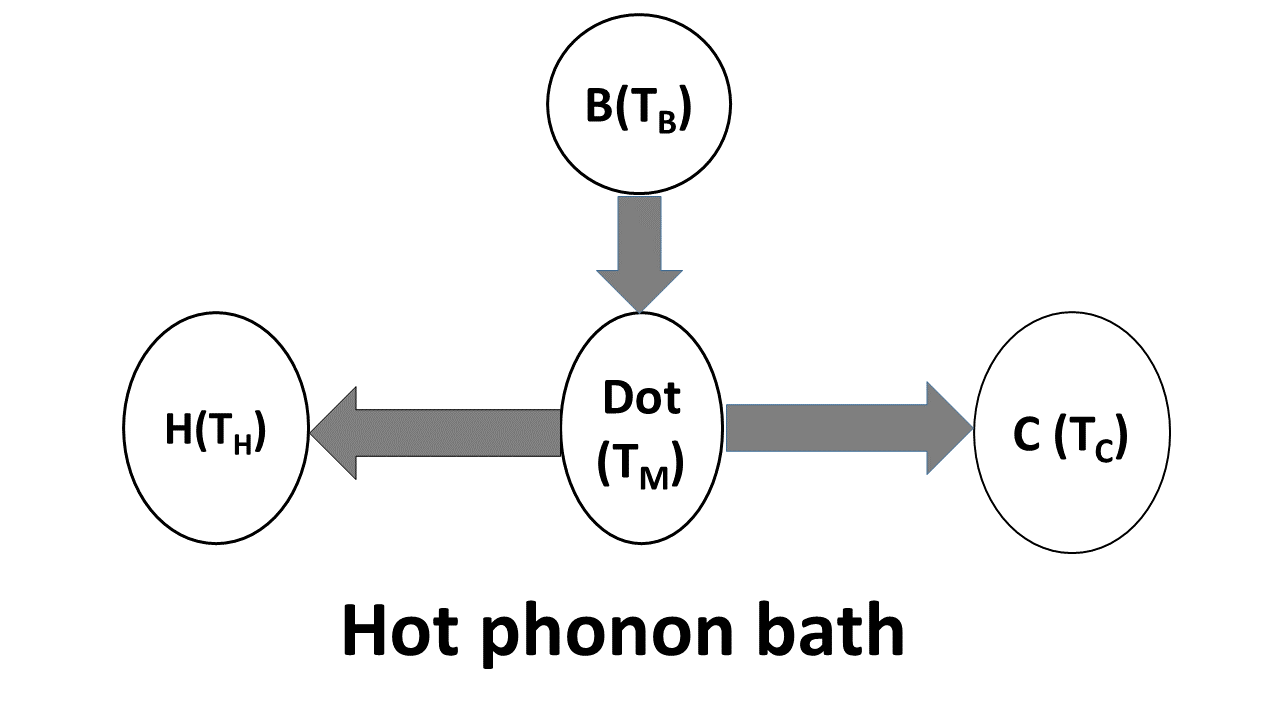}\label{7a}}
		\quad
		\subfigure[]{\includegraphics[height=0.22\textwidth,width=0.22\textwidth]{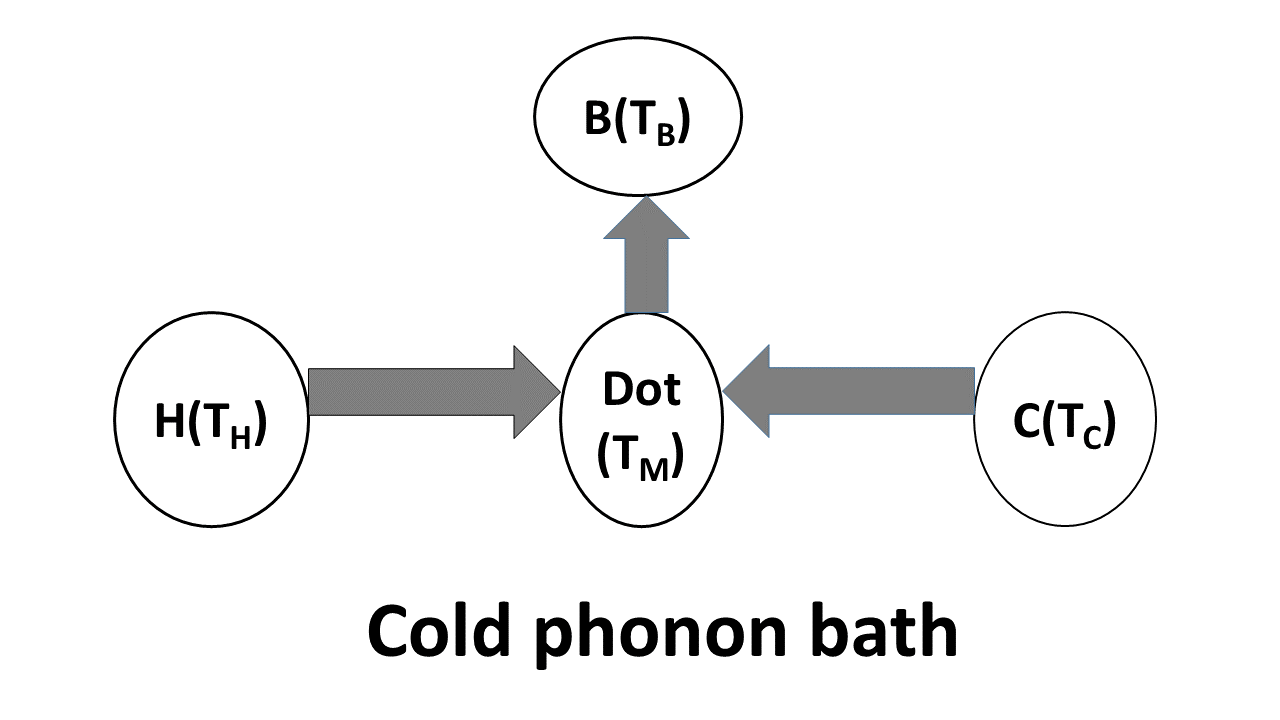}\label{7b}}
		\quad
		\subfigure[]{\includegraphics[height=0.221\textwidth,width=0.229\textwidth]{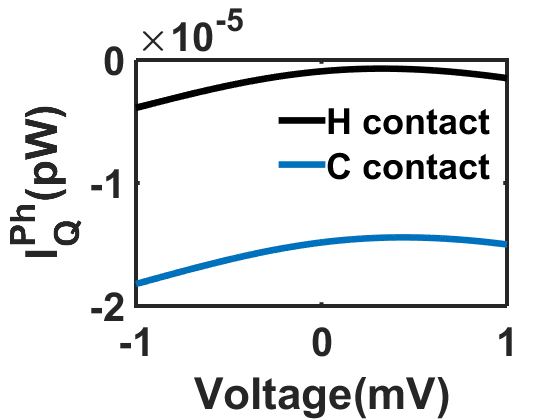}\label{7c}}
		\quad
		\subfigure[]{\includegraphics[height=0.221\textwidth,width=0.229\textwidth]{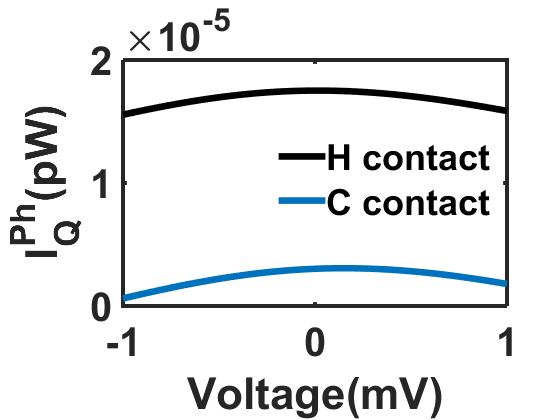}\label{7d}}
		\quad
		\caption{Evaluation of the sense of contact and bath phonon heat currents when the bath is strongly coupled to the dot for (a) Hot bath, $T_B$$>$$T_H$,$T_C$ and (b) Cold bath, $T_B$$<$$T_H$,$T_C$. (c),(d) Contact phonon heat currents for hot and cold bath respectively. (c) Note that the hot bath forces that the contact phonon currents flow away from the dot, and (d) that cold bath just reverses the direction.}
	\end{figure}
\end{center}
\subsection{Dot phonons coupled to a heat bath}
\indent In this section, we discuss exclusively the role of a heat bath in determining the thermoelectric performance of a heat engine and how the temperature of the bath controls the thermoelectric efficiency. In the limit of non-equilibrium phonons, the electronic heat current is much greater than the phonon heat current and takes the major role in determining $\eta$. But as the phonons of the dot become strongly coupled to the bulk phonon mode of any macroscopic body, the phonon heat current also becomes a relevant quantity. 
\indent We start by estimating the temperature of the quantum dot by coupling the central system with a thermometer phonon bath as described in earlier works \cite{Galperin2007, Galperin2006, Galperin2004}. This is based on the principle that the phonon heat current between the thermometer and the dot vanish when the temperature of thermometer equals the temperature of the dot, $T_M$. In Fig.~\ref{6a} and ~\ref{6b}, we present the trends of the molecular temperature as a function of $\gamma$/$\beta$ and applied voltage. We see that dot temperature, $T_M$, is a  weak function of the bias voltage. In Fig.~\ref{6c} and ~\ref{6d}, we have shown the dependence of $T_M$ with the electron-phonon coupling parameter $\lambda$. In the strong coupling limit, for $\lambda$=0, $T_M$ just follows $T_B$ and the bath phonon current cancels out. But as $\lambda$ increases, $T_M$ deviates from $T_B$ and gives rise to a bath phonon heat current.
\begin{center}
\begin{figure}[!htb]
		
		\subfigure[]{\includegraphics[height=0.221\textwidth,width=0.229\textwidth]{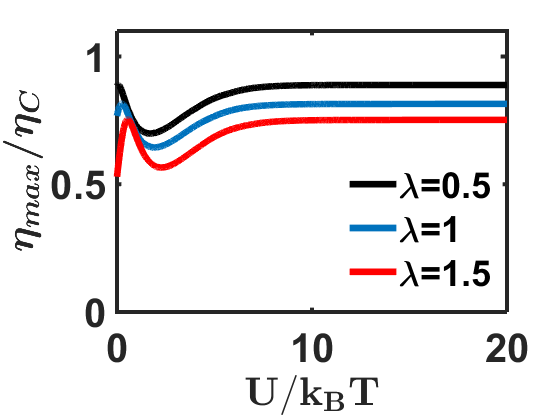}\label{8a}}
		\quad
		\subfigure[]{\includegraphics[height=0.221\textwidth,width=0.229\textwidth]{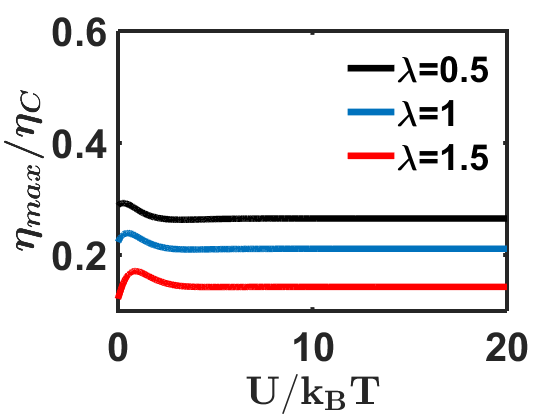}\label{8b}}
		\quad
		\caption{Variation of $\eta_{max}$ as a function of $U$ in the limit of strong coupling to the phonon bath. (a) Variation of efficiency as a function of $U$ when $T_B<T_H,T_C$ (b) Same plot for $T_B>T_H,T_C$. This figure establishes the fact that a cold bath does not affect $\eta$ much, while a hot bath does.}
			
		\end{figure}
		
	\end{center}
\indent To evaluate how $T_M$ controls the sense of contact phonon heat currents when the dot is strongly coupled to an external heat bath, We illustrate a schematic in Fig.~\ref{7a} and Fig.~\ref{7b}. For a non-zero $\lambda$, a hot bath always keeps $T_M$ greater than both $T_H$ and $T_C$, compelling the contact phonon heat currents to flow away from the dot. On the other hand, the hot bath itself pumps a phonon heat current into the dot. A cold bath does just the opposite and extracts phonon heat currents out of the dot which, in turn, compels phonon heat currents to flow from the contacts. In Fig.~\ref{7c} and Fig.~\ref{7d}, we show the plot of contact phonon heat currents considering that the dot is strongly coupled to the hot and the cold bath respectively. By convention, phonon currents from the contact to the dot are taken to be positive. We notice that for the hot bath, phonon heat currents flow away from the dot leading to a cooling of the dot by the contacts. A strongly coupled cold bath just does the reverse. If the bath is kept at an intermediate temperature, then the contact phonon heat currents will maintain the same direction, i.e., the hot contact will push phonons into the dot and the cold contact will extract phonons out of the dot. \\
\indent In Fig.~\ref{8a} and Fig.~\ref{8b}, we repeat the same plot as Fig.~\ref{4a} and Fig.~\ref{4b} for a fixed $\lambda$ as the temperature of the heat bath varies. We see that when the bath is hot ($T_B >>T_H, T_C$), the maximum efficiency reduces the most. The maximum efficiency $\eta_{M}$ improves monotonically as we reduce the bath temperature. The expression for the efficiency under the hot and cold bath conditions may be written as
	\begin{equation}
		\eta_{Hot}=\frac{P}{J_{el}^{Q}+J_{phB}^{Q}}
	\end{equation}
	\begin{equation}
		\eta_{cold}=\frac{P}{J_{el}^{Q}+J_{Q}^{phH}+J_{Q}^{phC}}.
	\end{equation}
	\begin{center}
		\begin{figure}[!htb]
			
			\subfigure[]{\includegraphics[height=0.218\textwidth,width=0.229\textwidth]{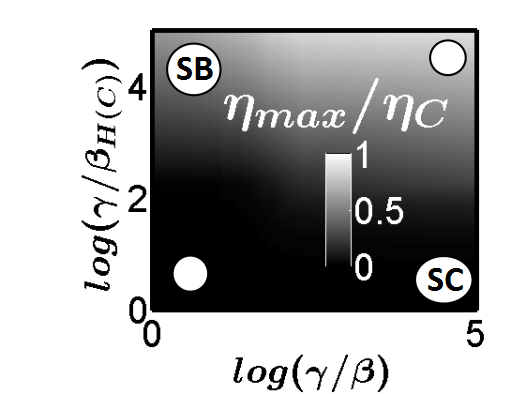}\label{9a}}
			\quad
			\subfigure[]{\includegraphics[height=0.218\textwidth,width=0.229\textwidth]{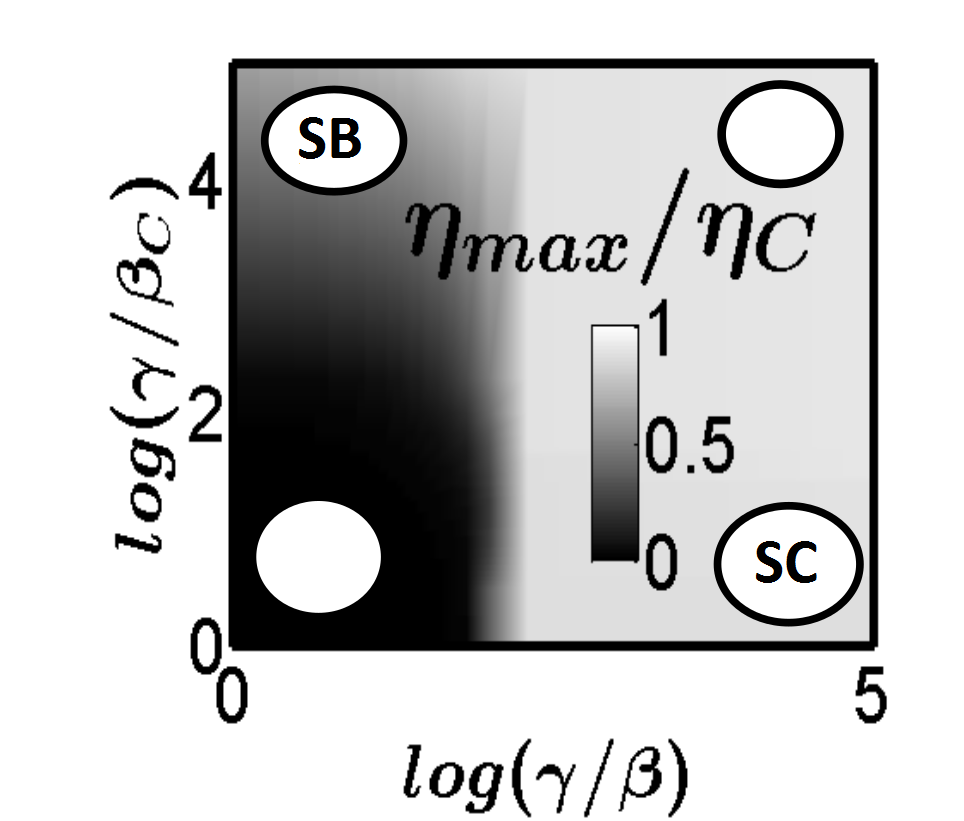}\label{9b}}
			\quad
			\subfigure[]{\includegraphics[height=0.221\textwidth,width=0.229\textwidth]{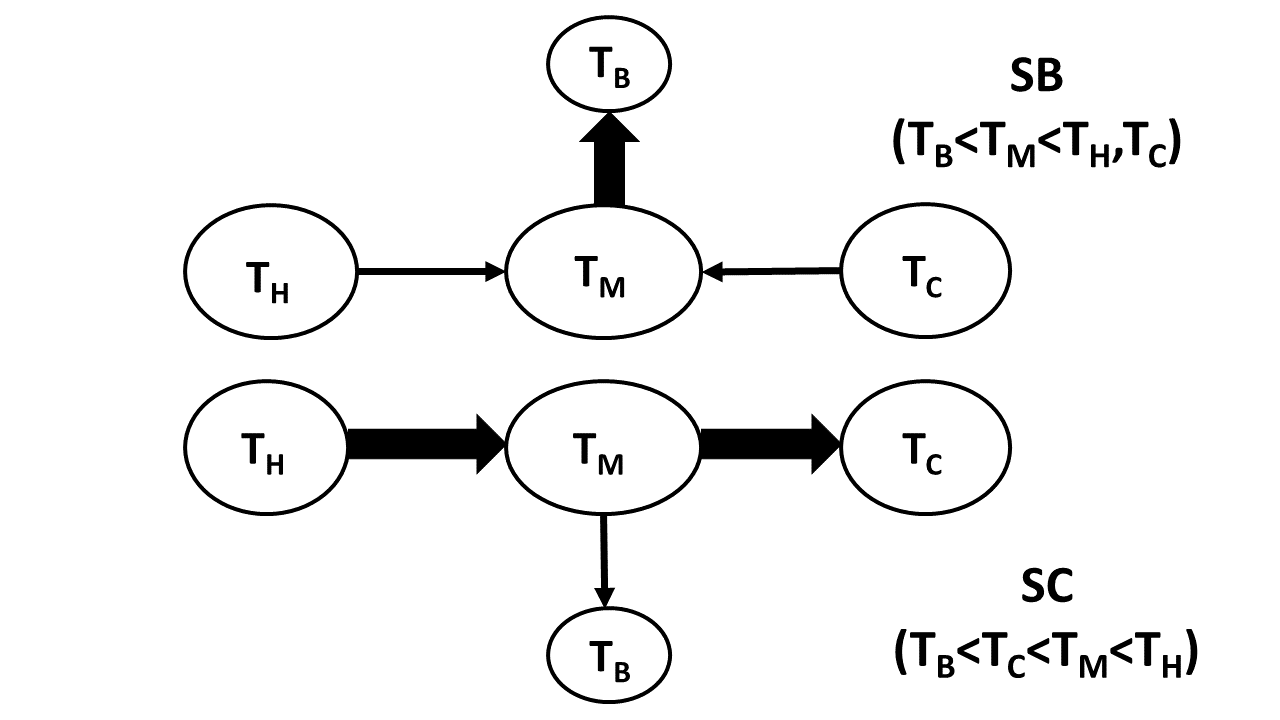}\label{9c}}
			\quad
			\subfigure[]{\includegraphics[height=0.221\textwidth,width=0.229\textwidth]{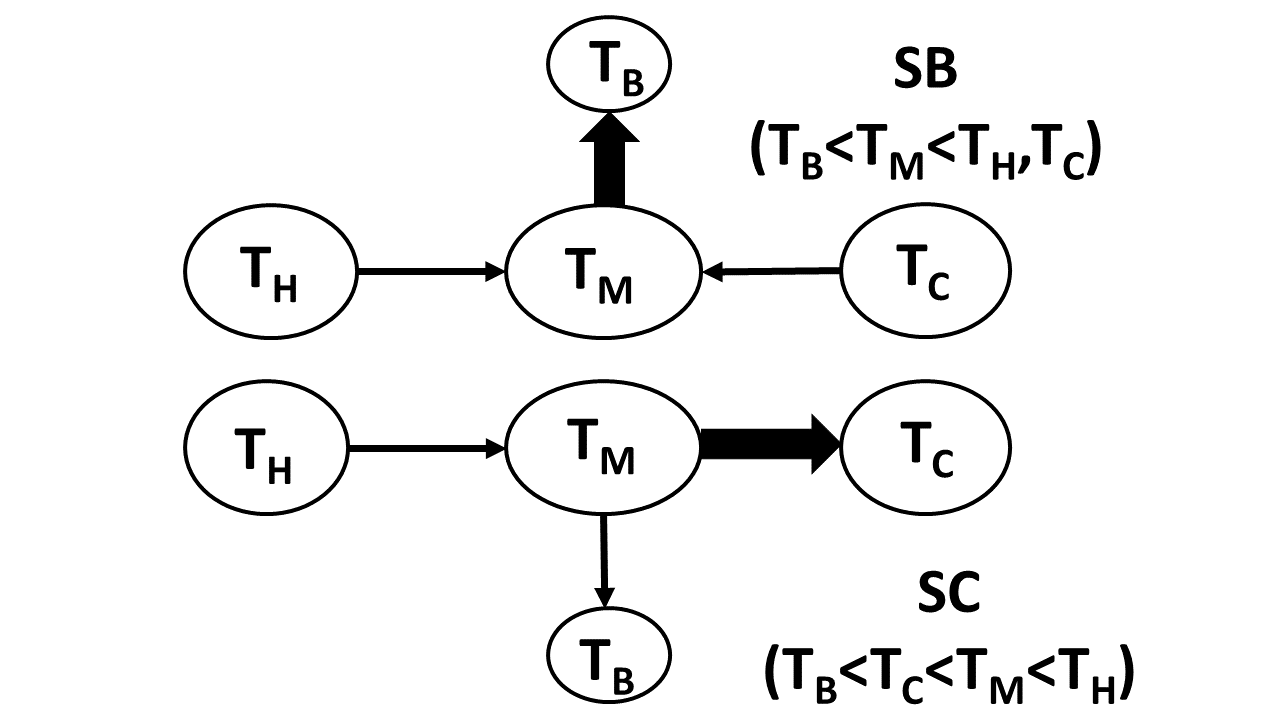}\label{9d}}
\caption{A trade-off between $\beta$ and $\beta_{H(C)}$ to optimize $\eta_{max}$. We consider that the heat bath is cold, i.e $T_B<T_H,T_C$. (a) 3-D plot showing the variation of $\eta_{max}$(normalized by $\eta_C$) as a function of $\beta$ and $\beta_{H(C)}$, when the dot is (a) symmetrically and (b) asymmetrically phonon coupled to the contacts. Two blank circles represent the regions which are very weakly and very strongly coupled to both the bath and the contacts. The region \textbf{SB} indicates where the dot is strongly coupled to the bath but weakly to the contacts. The region \textbf{SC} indicates the reverse case. (c) and (d), Schematic of the directions of the phonon heat currents in all cases considered. The thick (thin) arrows represent the sense of phonon heat currents between the dot and the strongly (weakly) coupled macroscopic bodies respectively. }
			
		\end{figure}
		
	\end{center}
The above expressions are based on the sense of the phonon heat currents. When the dot is strongly coupled to the hot bath, the input phonon heat current is supplied by the heat bath only. Whereas, when the dot is strongly coupled to a cold bath, the contacts supply phonon heat currents into the dot. Since $\beta>>\beta_{H},\beta_{C}$, a hot bath deteriorates $\eta$ much more in comparison with the cold bath. A heat bath with an intermediate temperature keeps $\eta$ in between. Thus, a quantum dot coupled to a cold environment ensures a better thermoelectric performance and merits a greater efficiency. 
\subsection{Trade-off between different phonon couplings and efficiency optimization }
 \indent In the earlier section, we have discussed the effect of a strong coupling to a heat bath and established that the efficiency, $\eta$, is controlled by the bath temperature. We must also note that the degree of phonon coupling between the dot and contacts is an important factor in determining $\eta_{max}$. In this section we investigate how $\eta_{max}$ is influenced as a function of $\beta$, $\beta_H$ and $\beta_C$ and present conditions on the optimization of $\eta_{max}$. The preceding section established that the efficiency can be improved by coupling the dot to a cold environment which drives the phonons out of the dot. Hence from now on we will focus on the situation where the heat bath is a cold one.\\	
\indent In Fig.~\ref{9a} we present the variation of $\eta_{max}$ as a function of the dot to bath and the dot to contact phonon couplings when $\beta_H=\beta_C$. When the dot is coupled strongly to both the contacts and the bath, $\eta_{max}$ is low. When the dot is weakly connected to both of them, phonons remain in non-equilibrium and hence this results in a high $\eta_{max}$. These two regions are represented by blank circles. In the regime strong coupling to the contact $\textbf{SC}$, $T_M$ remains close to the average temperature and the $H$ contact pushes large phonon currents to decrease $\eta_{max}$. On the other hand, in the regime of strong coupling to the bath, i.e., in the region marked as $\textbf{SB}$, $T_M$ remains close to the bath temperature and the cold bath extracts heat currents from the dot to increase $\eta_{max}$. The thermodynamics of phonon heat flow is shown in Fig.~\ref{9c}. Hence, when the dot is equally coupled to the contacts, strong bath coupling is better than strong contact coupling. \\
\indent  We now turn our attention to the case when the dot is asymmetrically coupled to both the contacts. In Fig.~\ref{9b}, we present a plot similar to that in Fig.~\ref{9a}. We are only interested in the case $\beta_H<<\beta_C$, where there is a chance of getting high $\eta_{max}$. Here we see just the opposite case. The $\textbf{SB}$ regime gives a similar performance like the earlier case. But $\eta_{max}$ is maximized in the regime $\textbf{SC}$ since phonon currents pushed in by the hot contact are smaller. Even in this case, we can ensure $\eta_{max}$ to be the same as the non-equilibrium case. Hence, this is the region where the efficiency is optimized.\\
\section{\label{sec:level4} Conclusion} This paper examined the thermoelectric response of a dissipative quantum dot heat engine based on the Anderson-Holstein model in two relevant operating limits (i) when the dot phonon modes are out of equilibrium, and (ii) when the dot phonon modes are strongly coupled to an external heat bath. In the first case, a detailed analysis of the related physics was elucidated and it was  conclusively demonstrated that an n-type heat engine performs better than a p-type as a result of an interplay between the on-site Coulomb interaction and the coupling to dot phonons. In the second case, with the aid of the dot temperature estimated by incorporating a {\it{thermometer bath}}, it was shown that the dot temperature deviates from the bath temperature as electron-phonon interaction becomes stronger. Consequently, we showed that the dot temperature intimately controls the direction of phonon heat current thereby influencing the thermoelectric performance. Our simulations highlight two crucial aspects: (a) a cold bath strongly coupled to the dot does not affect the efficiency that much but a hot bath does. (b) When the dot is phonon coupled with contacts, $H$ and $C$ and the cold bath $B$, it is better to couple it strongly to $B$ provided the phonon couplings with $H$ and $C$ are symmetric, whereas it is better to couple it strongly to $C$ if the phonon couplings with $H$ and $C$ are asymmetric. While the current work explored many aspects related to the functioning of a dissipative quantum dot heat engine, we believe some of the latter ideas developed here might merit a separate investigation by examining separately, the aspect of molecular Peltier cooling and refrigeration. \\ \\
{\it{Acknowledgements:}} Financial support from the Center of Excellence in Nanoelectronics (CEN) is acknowledged. We would like to thank Dr. S. D. Mahanti and Dr. R. H\"{a}rtle for illuminating discussions. 
	
\bibliographystyle{apsrev}

\bibliography{refrences}
\end{document}